\documentclass[12pt,a4paper]{article}

\usepackage{amsmath,amssymb}
\usepackage{graphicx}
\usepackage{hyperref}
\usepackage{cite}
\usepackage[margin=2.5cm]{geometry}

\usepackage{authblk}

\title{\bf Probing the CPT Constraint on CP Violation with $B^+ \to\pi^+ \mu^+ \mu^- $ Decay}
\author{Ignacio Bediaga
\thanks{bediaga@cern.ch}}

\affil{Centro Brasileiro de Pesquisas F\'isicas (CBPF), Rua Dr.\ Xavier Sigaud~150,
Rio de Janeiro, RJ 22290-180, Brazil}

\begin{document}

\maketitle

\begin{abstract}

The CPT theorem, combined with S-matrix unitarity, imposes nontrivial constraints
on CP asymmetries in B-meson decays ~\cite{Marshak1969}. A long-standing question is whether these constraints apply in an inclusive manner---globally across all final states---or exclusively, within a final state interaction (FSI) as argued by Wolfenstein ~\cite{Wolfenstein1991}. We present the theoretical framework underlying this
relationship between CP violation and the CPT theorem (CPV/CPT), deriving the
CP-violating rate differences through the S-matrix formalism with final-state
interactions (FSI). We propose that a precise measurement of the integrated CP
asymmetry in $B^+ \to \pi^+\mu^+\mu^-$ can unambiguously discriminate between
the two mechanisms.
\end{abstract}

\section{Introduction}

CP violation in B-meson decays is one of the central testing grounds of the Standard
Model and a window onto physics beyond it. The observation of large direct CP
asymmetries in charmless hadronic B decays has stimulated an extensive theoretical
program aimed at predicting individual decay rates and CP asymmetries~\cite{Bediaga2020,Bevan2014,LHCb2018}.

However, any consistent theory of CP violation must respect the constraints
imposed by fundamental symmetry principles. Among these, the CPT
theorem---which follows from Lorentz invariance, locality, and
unitarity---plays a special role. It guarantees that the total decay width of any
particle equals that of its antiparticle. Applied to the B-meson system, this
yields a sum rule over all partial widths that must vanish identically:
\begin{equation}
\sum_{f}\left[\Gamma(B \to f) - \Gamma(\bar{B} \to \bar{f})\right] = 0\,.
\label{eq:CPT_global}
\end{equation}

A stronger statement is found in the textbook treatment by Marshak, Riazuddin, and
Ryan~\cite{Marshak1969}. Their argument invokes not just CPT, but CPT together with
the unitarity of the S-matrix. Applying CPT to the amplitude for
a $B$ decay in a sample with same quantum numbers  $\{Q\}$ and using S-matrix unitarity within that block leads
to~\cite{Bediaga2014}
\begin{equation}
\sum_{n\in\{Q\}}\left[\Gamma(B \to n) - \Gamma(\bar{B} \to \bar{n})\right] = 0\,,
\label{eq:CPT_block}
\end{equation}
where the sum runs over all states in the block $\{Q\}$ of fixed quantum numbers.
This does not merely require the global sum to vanish; it requires each
quantum-number sector to be separately balanced. As an illustration of the strength
of this CPT constraint, consider the sum rule
$\Gamma(K^+ \to \pi^+\pi^-\pi^+) + \Gamma(K^+ \to \pi^+\pi^0\pi^0)
= \Gamma(K^- \to \pi^-\pi^-\pi^+) + \Gamma(K^- \to \pi^-\pi^0\pi^0)$ ~\cite{Marshak1969}.
The non-leptonic decay $K^\pm \to \pi^\pm\pi^0$ cannot enter this sum rule,
since G-parity forbids three pions from re-scattering into two. Another consequence of Eq.~(\ref{eq:CPT_block}), is the CP violation of the $K^\pm \to \pi^\pm\pi^0$ decay must be zero, due to the absence of a companion of this non-leptonic decay.

The large number of available channels for heavy-meson decays, particularly in B
decays, enlarges the set of final states sharing the same quantum numbers. For
example, B-meson decays without strangeness exchange ($\Delta S = 0$) can
encompass a large set of charmless B decays with two, three, or even more light
mesons, as well as decays involving charm--anti-charm mesons. The same holds for
$\Delta S = 1$ decays. Thus, Eq.~(\ref{eq:CPT_block}) apparently becomes a weak
constraint on CP asymmetries, which can be satisfied by balancing charmless and
charm decays with the same quantum numbers through S-matrix unitarity. We refer to
this as the \emph{inclusive mechanism} for satisfying the CPT constraint of
Eq.~(\ref{eq:CPT_block}).

Wolfenstein~\cite{Wolfenstein1991} proposed a more restrictive approach based on
Eq.~(\ref{eq:CPT_block}). His argument invokes not just CPT alone, but CPT in
conjunction with the unitarity of the S-matrix and the structure of final-state
interactions (FSI). We refer to this as the \emph{exclusive mechanism} for
satisfying the CPT constraint of Eq.~(\ref{eq:CPT_block}).

Within this framework, correlations between channels with complementary CP
asymmetries are generated through re-scattering---that is, through the
off-diagonal elements of the S-matrix. This overcomes the arbitrariness of the
inclusive mechanism, in which the CPT constraint is satisfied by uncorrelated CP
asymmetries arising from different dynamics.

The implication for charmless B decays is significant: the off-diagonal terms
have, in general, a strong-phase difference with respect to the diagonal ones.
As a consequence, these decays need not depend solely on charm-penguin loops,
which acquire an absorptive part only when the gluon virtuality
satisfies $q^2 > 4m_c^2$.

The recent LHCb measurement of CP asymmetry in $B^+ \to\pi^+ \mu^+ \mu^- $ \cite{LHCb2026FCNC} provides a new test of the two CPT constraint mechanisms discussed in this work. However, the LHCb measurement yielded an inconclusive, statistically limited result, but with good prospects for providing a definitive answer with Run 3 LHCb data.

\section{The \texorpdfstring{$CP$}{CP}-Violating Rate Difference}

Consider a particle $P$ decaying into a family with only two hadronic final
states, $\alpha$ and $\beta$, which interact with each other through re-scattering.
The diagrammatic contributions include:
\begin{itemize}
  \item $P \to \alpha$ (no re-scattering);
  \item $P \to \beta \to \alpha$ (single re-scattering);
  \item $P \to \beta $ (no re-scattering);
  \item $P \to \alpha \to \beta$ (single re-scattering).
\end{itemize}
In this scenario, the S-matrix is $2\times 2$ and takes the form:
\begin{equation}
S =
\begin{pmatrix}
e^{2i\delta_\alpha} & t_{\alpha\beta}\,e^{i(\delta_\alpha+\delta_\beta)} \\
t_{\alpha\beta}\,e^{i(\delta_\alpha+\delta_\beta)} & e^{2i\delta_\beta}
\end{pmatrix},
\label{eq:Smatrix2}
\end{equation}
where:
\begin{itemize}
  \item $\delta_\alpha$, $\delta_\beta$ are the elastic phase shifts of channels
        $\alpha$ and $\beta$, respectively;
  \item $t_{\alpha\beta}$ is the transition amplitude (inelasticity) between the
        two channels, satisfying $|t_{\alpha\beta}| \leq 1$.
\end{itemize}

\begin{figure}[h]
  \centering
\includegraphics[width=0.6\textwidth]{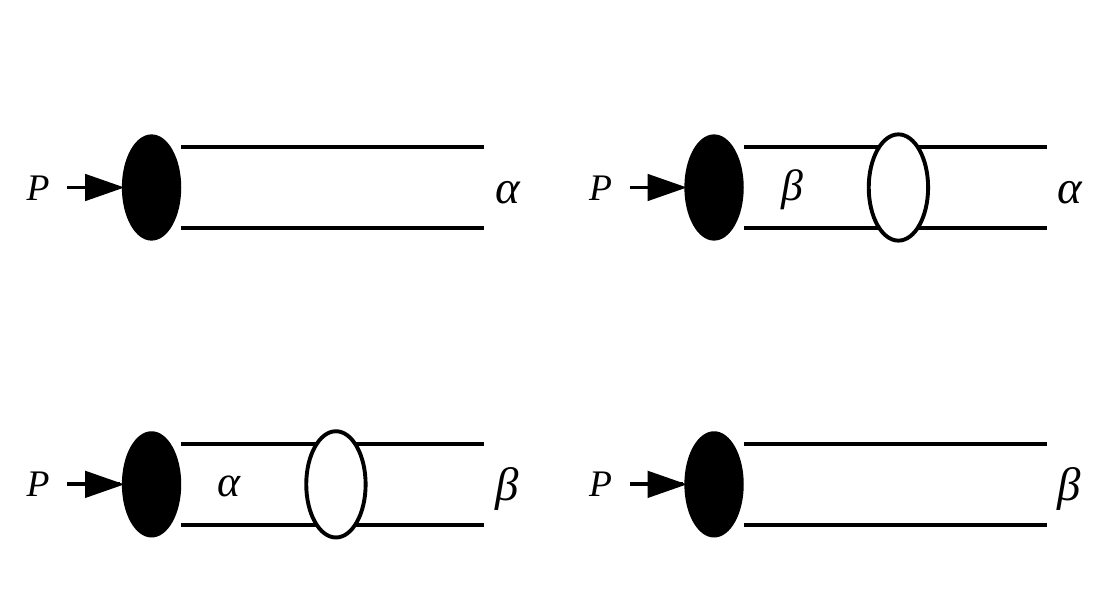}
  \caption{Graphical representation of  S-matrix Eq.~(\ref{eq:Smatrix2}).}
  \label{fig:re-scattering}
\end{figure}

The decay amplitudes $\langle i|T|P\rangle$ including final-state interactions
are:

\noindent\textbf{Antiparticle $\bar{P}$:}
\begin{align}
\langle \bar{\alpha}|T|\bar{P}\rangle &= e^{i\delta_\alpha}
  \bigl[T_\alpha + i\,t_{\alpha\beta}\,T_\beta\bigr], \label{eq:Pbar_a}\\
\langle \bar{\beta}|T|\bar{P}\rangle &= e^{i\delta_\beta}
  \bigl[T_\beta + i\,t_{\alpha\beta}\,T_\alpha\bigr]. \label{eq:Pbar_b}
\end{align}

\noindent\textbf{Particle $P$ (substitution $T_\rho \to T_\rho^*$):}
\begin{align}
\langle \alpha|T|P\rangle &= e^{i\delta_\alpha}
  \bigl[T_\alpha^* - i\,t_{\alpha\beta}\,T_\beta^*\bigr], \label{eq:P_a}\\
\langle \beta|T|P\rangle &= e^{i\delta_\beta}
  \bigl[T_\beta^* - i\,t_{\alpha\beta}\,T_\alpha^*\bigr], \label{eq:P_b}
\end{align}
where $T_\alpha$ and $T_\beta$ are the bare decay amplitudes in the absence of
FSI. The overall phase $e^{i\delta_i}$ in each channel is the statement of
Watson's theorem: the phase of the decay amplitude in channel $i$ equals the
elastic phase shift $\delta_i$ of that channel.

The difference in partial decay rates between antiparticle and particle for
channel $\alpha$ is:
\begin{equation}
\Delta_\alpha \equiv \Gamma(\bar{P} \to \bar{\alpha}) - \Gamma(P \to \alpha).
\label{eq:DeltaAlpha}
\end{equation}
The elastic phase $e^{i\delta_\alpha}$ has unit modulus and cancels in the
squared amplitude, so this results in:
\begin{equation}
\Delta_\alpha = \bigl|T_\alpha + i\,t_{\alpha\beta}\,T_\beta\bigr|^2
             - \bigl|T_\alpha^* + i\,t_{\alpha\beta}\,T_\beta^*\bigr|^2.
\label{eq:Delta_squared}
\end{equation}

Here $T_i \equiv r_i e^{i\phi_i}$, are the bare weak amplitudes carrying the CP-violating weak phase $\phi_i$.

Eq.~(\ref{eq:Delta_squared}) becomes for $\Delta_\alpha$ 
\begin{align}
\Delta_\alpha &= 4\,\mathrm{Im}(T_\alpha^*\,T_\beta)\,t_{\alpha\beta}\,.
\label{eq:Delta_alpha}
\end{align}
The same procedure can be applied to $\Delta_\beta$
\begin{align}
\Delta_\beta  &= 4\,\mathrm{Im}(T_\beta^*\,T_\alpha)\,t_{\alpha\beta}\,.
\label{eq:Delta_beta}
\end{align}
Since $\mathrm{Im}(T_\beta^*\,T_\alpha) = -\mathrm{Im}(T_\alpha^*\,T_\beta)$,
we confirm that $\Delta_\beta = -\Delta_\alpha$, consistent with the CPT
constraint that the sum must vanish. These equations show that the CP asymmetry is determined solely by the off-diagonal elements of the S-matrix. This is the central result of Wolfenstein's paper~\cite{Wolfenstein1991}.

The re-scattering process $\pi^-\pi^+ \to K^-K^+$ is straightforwardly
represented in this formalism with $\alpha = \pi^-\pi^+$ and
$\beta = K^-K^+$, and has been widely applied together with the CPT constraint
of Eq.~(\ref{eq:CPT_block}) in CP-violation studies of charmless three-body B
decays, including $B^\pm \to \pi^\pm\pi^-\pi^+$, $B^\pm \to K^\pm K^-K^+$,
$B^\pm \to \pi^\pm K^-K^+$, and $B^\pm \to K^\pm\pi^-\pi^+$. These studies
exploit the well-known $\pi^-\pi^+ \to K^-K^+$ re-scattering \cite{Pelaez2018}to relate the CP
asymmetries of the $\Delta S = 0$ modes $B^\pm \to \pi^\pm\pi^-\pi^+$ and
$B^\pm \to \pi^\pm K^-K^+$ to those of the $\Delta S = 1$ modes
$B^\pm \to K^\pm K^-K^+$ and $B^\pm \to K^\pm\pi^-\pi^+$~\cite{Bediaga2014,Nogueira2015,Garrote2023}.
These results were restricted to the invariant-mass region where
$\pi^-\pi^+ \to K^-K^+$ re-scattering is active, between 1 and 1.5~GeV.

\subsection{Generalized S-Matrix}

Generalizing the S-Matrix to n decay channels, this takes the form: let $\{|1\rangle, |2\rangle, \ldots, |n\rangle\}$ be a set of $n$ hadronic
states that can be produced in the decay of $P$ and that interact with one
another through re-scattering. The S-matrix is then an $n \times n$ operator
acting on this channel space. Let $\delta_i \in \mathbb{R}$ be the elastic
phase shift of channel $i$ and $t_{ij} = t_{ji} \in \mathbb{C}$ be the
transition amplitude between channels $i$ and $j$.

The complete matrix is:
\begin{equation}
S =
\begin{pmatrix}
e^{2i\delta_1}              & t_{12}\,e^{i(\delta_1+\delta_2)} & t_{13}\,e^{i(\delta_1+\delta_3)} & \cdots & t_{1n}\,e^{i(\delta_1+\delta_n)} \\
t_{12}\,e^{i(\delta_1+\delta_2)} & e^{2i\delta_2}              & t_{23}\,e^{i(\delta_2+\delta_3)} & \cdots & t_{2n}\,e^{i(\delta_2+\delta_n)} \\
t_{13}\,e^{i(\delta_1+\delta_3)} & t_{23}\,e^{i(\delta_2+\delta_3)} & e^{2i\delta_3}              & \cdots & t_{3n}\,e^{i(\delta_3+\delta_n)} \\
\vdots                      & \vdots                      & \vdots                      & \ddots & \vdots \\
t_{1n}\,e^{i(\delta_1+\delta_n)} & t_{2n}\,e^{i(\delta_2+\delta_n)} & t_{3n}\,e^{i(\delta_3+\delta_n)} & \cdots & e^{2i\delta_n}
\end{pmatrix}.
\label{eq:Smatrix_n}
\end{equation}
Let $T_i^{(0)}$ be the bare decay amplitude $P \to i$ (in the absence of
final-state interactions). Including the FSI described by the S-matrix, the
physical decay amplitude into channel $i$ is~\cite{Wolfenstein1991,Bigi2009}:
\begin{equation}
\boxed{
\langle i\,|\,T\,|\,P\rangle = e^{i\delta_i}
\left[T_i^{(0)} + i\sum_{j\neq i} t_{ij}\,T_j^{(0)}\right].
}
\label{eq:FSI_amplitude}
\end{equation}
Using the same procedure as in the two-channel example, we arrive at the
generalized equations for $n$ channels:
\begin{align}
\Delta_\alpha &= 4\sum_{\beta\neq \alpha}\mathrm{Im}(T_\alpha^*\,T_\beta)\,t_{\alpha\beta}\,,
\label{eq:Delta_alpha_gen}\\
\Delta_\beta  &= 4\sum_{\alpha \neq \beta}\mathrm{Im}(T_\beta^*\,T_\alpha)\,t_{\alpha\beta}\,.
\label{eq:Delta_beta_gen}
\end{align}
The generalized result for $n$ scattering final states is the same as that
obtained above for only two channels: $\Delta_\beta = -\Delta_\alpha$,
satisfying Eq.~(\ref{eq:CPT_block}).

The transition amplitude $t_{ij}$ with $i \neq j$ between channels in the
S-matrix~(\ref{eq:Smatrix_n}) encompasses all available re-scattering channels
with the same quantum numbers. As pointed out in the introduction, this includes
charmless B decays with two, three, or even more light mesons, as well as decays
involving charm--anti-charm mesons.

In the absence of a first-principles theory of hadronic interactions, it becomes
very difficult to determine the various $t_{ij}$. Moreover, the off-diagonal
elements are not constant parameters that hold across all available phase space;
they depend strongly on the center-of-mass energy. The two-body re-scattering
example discussed above, $\pi^-\pi^+ \to K^-K^+$, is significant only between 1
and 1.5~GeV~\cite{Pelaez2018}. Above this value, this re-scattering amplitude
becomes negligible.

There is abundant experimental information on hadron--hadron scattering at low
invariant masses. In the region below 2~GeV, measured amplitudes and strong-phase
differences are available for $\pi\pi$, $K\pi$, and $KK$ scattering, including
elastic scattering and the light-meson resonance spectrum. These experimental
results can serve as inputs for studies of a large sample of light-meson
multi-body B decays, since these decays are dominated by two- or three-body
low-mass resonances. Several charmless three-body B-decay amplitude analyses were
performed using this experimental information~\cite{LHCb2026,LHCb_3pi_2019,LHCb_kkpi_2018}.

On the other hand, no analogous experimental information is available for
charmless non-leptonic two-body B decays. At 5~GeV, well above the light
resonance region, the $\pi\pi$, $K\pi$, and $KK$ scattering amplitudes are not
well determined, and these final states can receive contributions from
intermediate double-charm decays, which themselves contribute importantly to B
decays.

\section{$B^+ \to \pi^+\mu^+\mu^-$ decay as probe}

The overall decay $b \to d\ell^+\ell^-$ receives comparable contributions from
three CKM combinations: $V_{tb}V_{td}^*$, $V_{cb}V_{cd}^*$, and $V_{ub}V_{ud}^*$.
Each carries a different weak phase from the CKM matrix and acquires a different
strong phase through $c\bar{c}$ and $u\bar{u}$ quark loops. The coexistence of
these different phases produces a CP asymmetry that varies with the $\ell^+\ell^-$
invariant mass. Following Kr\"uger and Sehgal~\cite{Kruger1997a}, this asymmetry
can reach 25\%, depending on $m_{\ell^+\ell^-}$. Beyond this $b \to d\ell^+\ell^-$
prediction, the possible presence of CP violation was also studied by other
authors, who predicted significant CP asymmetries in the decay
$B^+ \to \pi^+\mu^+\mu^-$ as a function of the $\mu^+\mu^-$ invariant
mass~\cite{Kruger1997b,Du2015,Hurth2017,Biswas2023,Becirevic2024}. All these
studies are based on the inclusive mechanism for satisfying the CPT constraint.

However, if one imposes the exclusive mechanism to satisfy CPT, as proposed by
Wolfenstein, the decay $B^+ \to \pi^+\mu^+\mu^-$ constitutes a single-element
re-scattering block, analogous to $K^+ \to \pi^+\pi^0$, so the total integrated
CP asymmetry must vanish. The CPT constraint applies only to the integrated total
decay width; the differential asymmetry $d\mathcal{A}_{CP}/dq^2$ can be non-zero
at individual values of $q^2$, provided it averages to zero over the full
spectrum.

The existence of decay channels with negligible CP asymmetry, despite the
simultaneous presence of different weak and strong phases, has been discussed in
the literature in the context of the exclusive mechanism for satisfying CPT. It
was examined for charmless three-body B decays involving an intermediate vector
resonance~\cite{Nogueira2016}. Experimental results partially agree with this
proposal~\cite{LHCb2023}, with the exception of the $B^+ \to K^+\rho^0$
decay~\cite{LHCb:2026CP}.

Charmless three-body B decays have proven to be a rich testing ground for
understanding properties of CP-violation dynamics. However, the influence of the
spectator particle in the two-body interaction is not clearly established. The $2+1$ approach is a widely used approximation in all amplitude-analysis methods.
Consequently, the relative merits of the inclusive and exclusive mechanisms for
satisfying CPT through Eq.~(\ref{eq:CPT_block}) remain controversial.

\begin{figure}[h]
  \centering
  \includegraphics[width=0.65\textwidth]{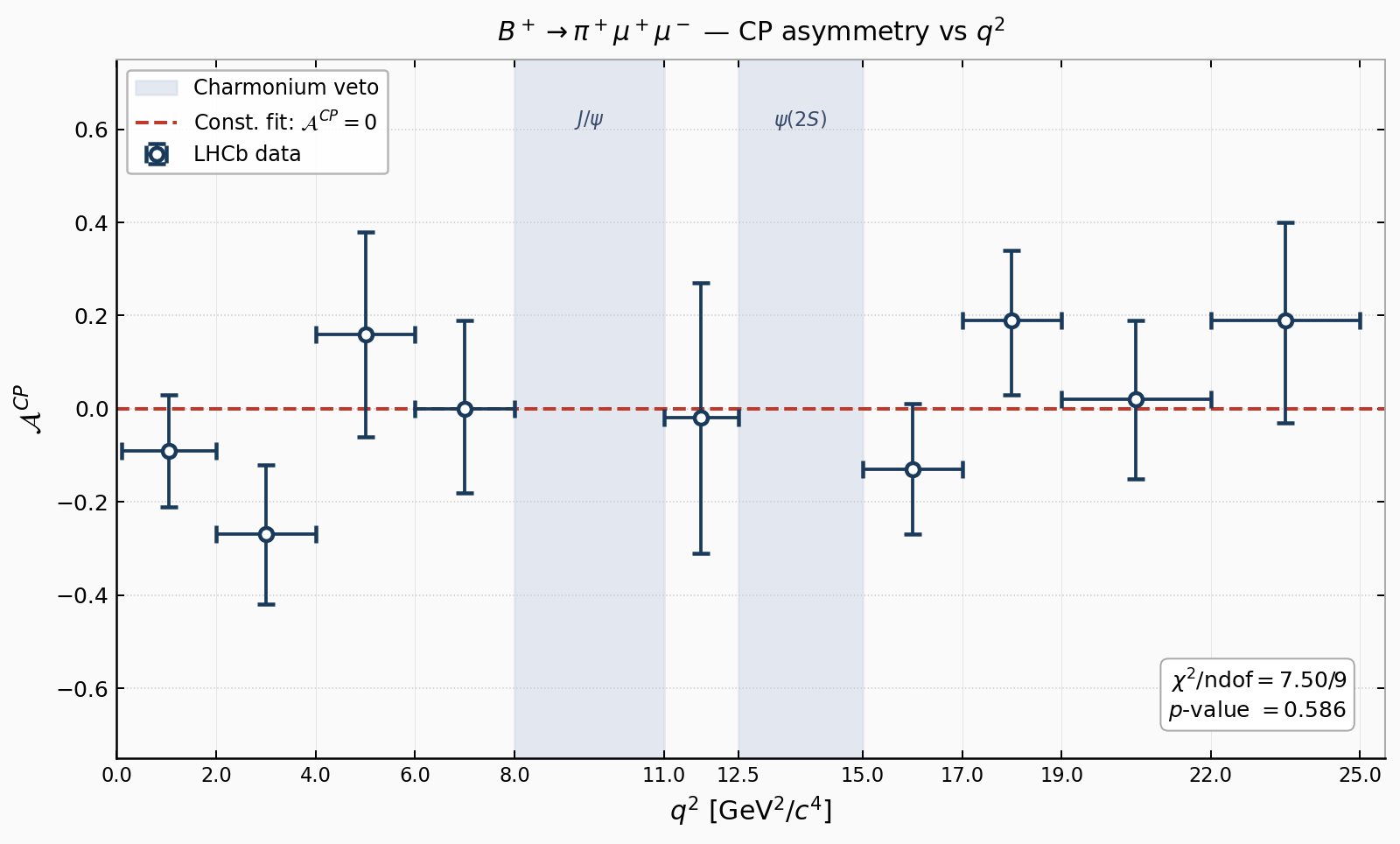}
  \caption{CP asymmetry distribution for $B^+ \to \pi^+\mu^+\mu^-$, evaluated
    in bins of the dimuon invariant-mass squared~\cite{LHCb2026FCNC}. The fit
    function is a constant representing zero CP-violation
    hypothesis.}
  \label{fig:ACP_Bpi}
\end{figure}

The experimental study of $B^+ \to \pi^+\mu^+\mu^-$ could resolve this
controversy. The LHCb collaboration recently presented an experimental
study of this channel, specifically searching for CP asymmetry~\cite{LHCb2026FCNC}.
The measurements were performed in bins of the dimuon invariant-mass squared,
using an integrated luminosity of $9\,\mathrm{fb}^{-1}$ recorded during Runs~1
and~2 of the LHCb experiment. The result is presented in Fig.~\ref{fig:ACP_Bpi}.

The limited statistical precision does not allow a conclusive discrimination
between the inclusive and exclusive approaches to the CPT constraint based on the
observed CP average value. The constant function fixed in zero is well compatible with the zero CP-violation hypothesis yielding a good $\chi^2= 7.5/9$ and a high $p = 0.586$-value. Other than the total integrated value compatible with $A_{CP}=0$, 
an important factor supporting the exclusive approach is the good performance of the constant fit along $q^2$. In fact, the general predictions of
Refs.~\cite{Kruger1997b,Du2015,Hurth2017,Biswas2023,Becirevic2024} are:
(i)~the sign of the CP asymmetry is predominantly negative ($B^-$ decays more frequently
than $B^+$ in the dimuon mass spectrum); and (ii)~the asymmetry is largest at
low $q^2$ (high recoil, low dimuon mass) and decreases towards the $J/\psi$.
Neither feature appears clearly in Fig.~\ref{fig:ACP_Bpi}.

The recently completed LHCb Run~3 data-taking campaign is expected to increase
the available dataset by a factor of approximately three relative to the present
result. We believe this will provide sufficient statistical power to decide
between the two approaches to the CPT constraint for CP asymmetry in FCNC decays,
and to shed decisive light on the dynamical mechanism underlying this asymmetry,
including the non-leptonic decays of charm and beauty particles.

\section{Multi-Hadron FCNC \texorpdfstring{$B$}{B} Decays}

To satisfy the exclusive CPT constraint while simultaneously observing a non-zero
CP asymmetry in FCNC B decays, one must consider final states containing two or
more hadrons. Even with two hadrons in the final state, restrictions apply. At low invariant mass (less than 1 $GeV$), a two-pion S-wave can rescatter only into another two-pion
S-wave state; the same holds for higher partial waves. Moreover, while two pions
can in principle rescatter into four pions, this process is not simply a
phase-space effect: elastic $\pi^+\pi^-$ scattering opens the four-pion channel
only above 1.6~GeV. In contrast, $\pi^+\pi^- \to K^+K^-$ re-scattering is
sizeable from near 1~GeV up to 1.5~GeV, and then closes~\cite{Pelaez2018}.

Taking into account the hypothesis of the dominance of the exclusive CPT
constraint in FCNC decays strongly motivates searching for CP violation in FCNC
B decays with two or more hadrons in the final state. In this framework,
multi-hadron final states provide the off-diagonal S-matrix elements necessary
to generate, sustain, and redistribute CP asymmetries, while also supplying the
natural strong phase required to produce them.

To conclude, the question of whether the CPT sum rule is satisfied through an inclusive or an exclusive mechanism has concrete experimental consequences. We have shown that the decay $B^+ \to \pi^+\mu^+\mu^-$ is ideally suited to resolve this ambiguity: under the inclusive mechanism, significant CP asymmetries varying with the dimuon invariant mass are expected, while the exclusive (Wolfenstein) mechanism predicts
a vanishing integrated asymmetry, since this channel forms a single-element block
with no hadronic re-scattering partner. Existing LHCb measurements using
$9\,\text{fb}^{-1}$ from Runs~1 and~2 are consistent with zero asymmetry,
offering preliminary support for the exclusive mechanism, though the statistical
precision is not yet sufficient for a definitive conclusion. The Run~3 dataset,
expected to be roughly three times larger, should provide the discriminating power
needed to settle this fundamental question.

\section*{Acknowledgments}
The author thanks Felipe Luan, Jussara Miranda, and Rafael Coutinho
for enlightening discussions and for their careful reading of the manuscript
and valuable comments.

\end{document}